\documentclass[11pt,twoside]{article}

\usepackage{asp2006}
\usepackage{epsf}
\usepackage{psfig}
\usepackage{lscape}

\begin{document}
\title{Late stages of stellar evolution and their impact on spectrophotometric 
properties of galaxies}   
\author{Alberto Buzzoni}   
\affil{INAF - Osservatorio Astronomico,\\
Via Ranzani 1, 40127 Bologna (Italy)}    

\begin{abstract} 
The connection between AGB evolution of stellar populations and infrared vs. 
ultraviolet properties of the parent galaxies is reviewed relying on the 
updated lookout provided by population-synthesis theory. In particular, 
planetary-nebula events and hot horizontal-branch evolution are assessed in 
a unitary view to outline a plain general picture of galaxy spectrophotometric 
evolution. This will include a brief discussion of relevant phenomena such as 
the ``UV upturn'' in ellipticals and the stellar mass loss properties along 
the galaxy morphological sequence. 
\end{abstract}

\section{Introduction}

A detailed and physically self-consistent modelling of post Main Sequence (MS)
stellar evolution has been a challenging effort of theoretical astrophysics since the 
decade of the sixties \citep[see, e.g.][and references therein for a documental 
overview of the relevant pioneering works since then]{iben67}.
One main issue in this regard is that post-MS 
lifetime of stars (at every mass range) is no longer governed by the nuclear timescale 
alone, but different mechanisms intervene to strongly affect stellar structure, 
acting from ``inside out'' (e.g.\ convection) and from ``outside in'' (e.g.\ mass 
loss by stellar wind).

This forcedly restrains any theoretical output to a preliminary empirical validation process 
by matching real observations of nearby resolved stellar systems in order to suitably tune 
up the wide range of free parameters potentially allowed by theory.\footnote{The study of 
Galactic globular cluster c-m diagrams is an illuminating example in this sense 
\citep[e.g.][]{rffp,chiosi}.}
 Any successful approach in this sense, however, suffers from evident limitations
as far as distant (unresolved) galaxies are taken into account in our analysis.
By surveying different cosmic epochs and environment conditions, in fact, one might 
need to leave apart the local interpretative framework, typically constrained by 
the observation of low-mass metal-poor stars.

To overcome this potentially shifty bias in the analysis of deep cosmological data, and
considering that post-MS stars alone provide typically over 2/3 of galaxy total luminosity 
in bolometric \citep{buzzoni95,buzzoni05}, it is of paramount importance to assess on a more 
firm basis the leading phenomena that constrain stellar evolution along its latest stages, like 
the horizontal (HB) and the asymptotic (AGB) giant branches, among galaxies of different 
morphological type.

\section{Horizontal branch morphology and the ``UV-upturn'' phenomenon in elliptical galaxies}

The so-called ``UV-upturn'', that is the rising
ultraviolet emission shortward of 2000 \AA, sometimes featuring the 
spectral energy distribution (SED) of elliptical galaxies and the bulges of spirals
\citep{code79} has been for long a puzzling problem for such old galaxy environments dominated by 
stars of mass comparable to the Sun. 

\begin{figure}
\centerline{
\psfig{file=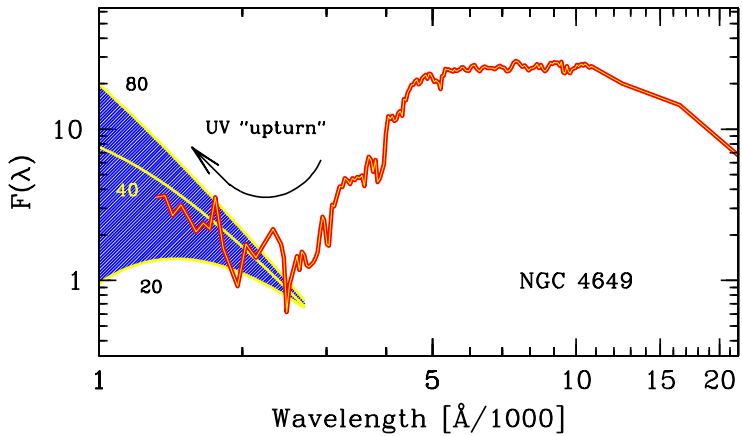,width=0.78\hsize,clip=}
}
\caption{The observed SED of the Virgo elliptical galaxy NGC~4649, one of the best 
example of the ``UV upturn'' phenomenon. The ultraviolet raising branch, shortward of
2000~\AA is matched with three black-body curves, for 20, 40 and 80,000~K, as labelled
on the plot. It is evident that stars around 40,000~K should be the main contributors
to the observed UV galaxy emission, supplying in total about 2\% of the 
galaxy bolometric luminosity.}
\label{n4649}
\end{figure}

In fact, the implied existence of an important contribution of (long-lived) 
O-B stars, hotter than 30\,000 - 40\,000~K and providing in the most striking cases
about 2\% of the galaxy bolometric luminosity (see Fig.~\ref{n4649}), has been sometimes 
identified with binaries, blue stragglers, blue HB stars, AGB {\it manqu\'e} stars, and 
post-AGB nuclei of planetary nebulae (PNe) \citep[see][for an updated review on this subject]{yiyoon04}.
Spectroscopy \citep{brown97} and imaging \citep{brown00} of resolved c-m diagrams for 
stellar populations in local galaxies, like M32, have definitely established that this UV excess 
mostly arises from the hot tail of a broad temperature distribution of HB stars further 
complemented, to a lesser extent, by a PN contribution. 

\begin{figure}[!t]
\centerline{
\psfig{file=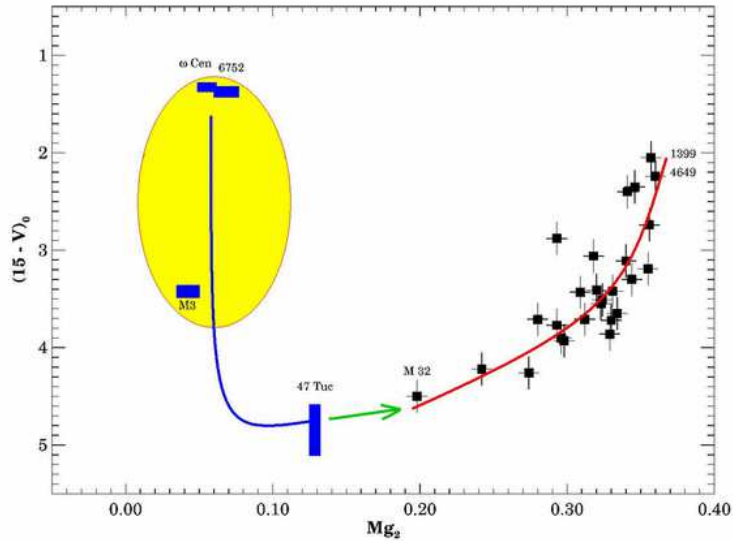,width=0.78\hsize,clip=}
}
\label{oconnell}
\caption{An illustrative sketch comparing UV-to-optical emission of Galactic globular
clusters (vertical ellipse to the left) and elliptical galaxies (square-marker sequence
to the right of the plot) after \citet{oconnell99}. The UV color is defined as 
$(15-V) = -2.5\,\log [F(1500{\rm \AA})/F(V)]$. Some relevant objects are labelled for both groups.
The Mg$_2$ Lick index is taken as standard metallicity indicator on the x-axis.
It is evident that UV-enhanced integrated colors can be reached at the two extremes of the
metallicity scale, in case of a blue HB morphology for metal-poorer globulars and due to 
EHB stellar contribution in case of the most metal-rich (giant) ellipticals. See text for
discussion.
}
\end{figure}

As a blue HB morphology is more comfortably produced
in old metal-poor globular clusters \citep[e.g.][]{rood73} then, on this line, one
should admit that UV stars in ellipticals represent the $Z \ll Z_\odot$ tail of an 
ostensibly broad metallicity distribution peaked at much higher values, around the solar 
abundance.\footnote{A metal-rich chemical composition should be advocated for the star bulk 
in ellipticals given, for instance, a much stronger integrated Mg$_2$ Lick index for 
these galaxies, compared to Galactic globular clusters 
\citep[see Fig.~\ref{oconnell}, and an extensive discussion in][]{oconnell99}.}

\begin{figure}[!t]
\centerline{
\psfig{file=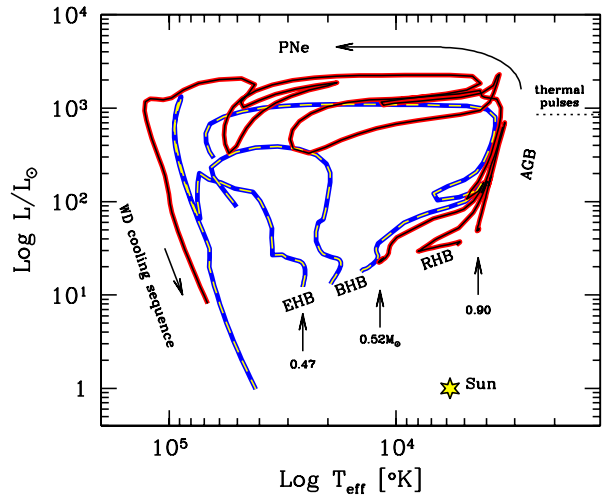,width=0.64\hsize,clip=}
}
\caption{A selected set of HB evolutionary tracks with $Z = Z_\odot$ from \citet{dorman}. 
The starting track envelope roughly identify the HB locus with the value of three relevant 
stellar masses (namely 0.47, 0.52 and 0.90~M$_\odot$) labelled on the plot.
Note the increasingly hotter HB temperature with decreasing stellar mass, with stars
below $\sim 0.50$~M$_\odot$ occupying the EHB region of the diagram, pertinent to SdO/SdB spectral
types, fully escaping the AGB phase (these are the so-called {\it ``AGB-manqu\'e''} stars)
and directly fading along the white-dwarf (WD) cooling sequence. 
The reported value of $\sim 0.52$~M$_\odot$ is the threshold for post-HB stars to end up as PNe 
after completing AGB evolution and undergoing the thermal pulsing phase.
The position of the Sun on the plot is located as a general reference for the reader.
}
\label{hbmorph}
\end{figure}

On the other hand, hot metal-rich HB stars might also be naturally predicted providing 
stars to approach the HB phase with a conveniently low external envelope, compared 
to their inner Helium core mass \citep{dorman,dcruz96}.
Figure~\ref{hbmorph} is an illustrative example in this sense, displaying a full set of 
HB models of solar metallicity (i.e.\ red, intermediate and very blue stars) and their 
involved post-HB evolutionary paths, based on the \citet{dorman} work. So-called ``Extreme HB'' 
stars (EHB), to be associated with hot SdO/SdB spectral types, have 
actually been observed, for example in $\omega$~Cen \citep{dcruz00} and in some old Galactic 
open clusters as well, like NGC~6791 \citep{kaluzny92,buson06}.

Though supplying a straightforward evolutionary framework for UV-en\-hanced elliptical galaxies, 
this hypothesis implies a quite delicate tuning of core mass size at the HB onset
(as a result of RGB nuclear burning processes) and mass loss efficiency (to suitably ``peel off'',
at the same time, stellar envelope). As a consequence, one has to expect the UV-to-optical color to 
be a quite fragile and quickly evolving feature in the SED of elliptical galaxies \citep{park97}.

\begin{figure}[!t]
\centerline{
\psfig{file=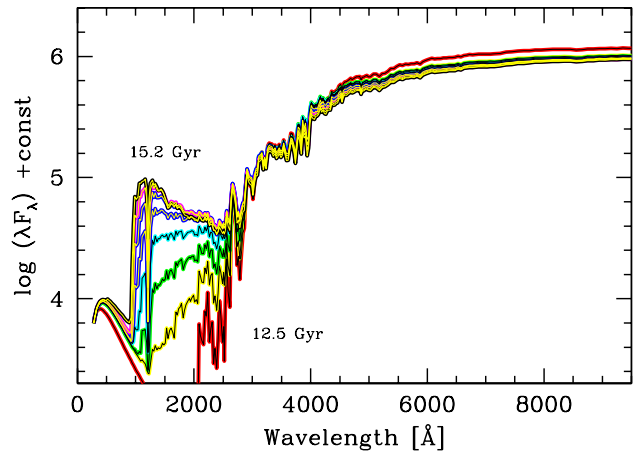,width=0.68\hsize,clip=}
}
\caption{Integrated SED for a SSP of solar metallicity, Salpeter IMF and \citet{reimers} mass-loss
parameter $\eta = 0.5$ after \citet{buzzoni89}. Spectral evolution is computed at steps of 200~Myr 
from $t = 15.2$ to 14~Gyr, plus a further model at 12.5~Gyr, as labelled on the plot. A broad 
temperature distribution of HB stars (matching the observed M3 HB morphology) is assumed at 
$t = 15.0$~Gyr. Note the quick evolution of the UV excess between 1000 and 2500~\AA, that 
already disappeared in the 12.5~Gyr model in consequence of the corresponding reddening of the 
HB color distribution. A second and nearly steady minor UV bump can be recognized  
shortward of 1000~\AA\ due to the contribution of PN nuclei.}
\label{uvupturn}
\end{figure}

This is confirmed in Fig.~\ref{uvupturn}, where we track back-in-time evolution of a 15 Gyr simple 
stellar population (SSP) of solar metallicity and intermediate HB morphology
(such as to closely resemble the temperature distribution of stars in the M3 globular
cluster), according to \citet{buzzoni89} population synthesis code.
Due to the presence of stars of increasingly higher mass at early epochs (giving rise to a red
HB), one sees from the figure that the full UV burst event quickly recovers in about 3 Gyr, that 
is barely a $\sim 20$~\% of galaxy's entire life. The UV-upturn can therefore fade by several magnitudes 
as the lookback time increases by a few Gyrs, making the effect in principle detectable at 
intermediate redshift ($z = 0.2-0.3$) \citep[][see also Ree et al, this conference]{brown03}.

\section{Planetary Nebulae and the Initial-to-Final mass relation}

Along the SSP evolution, a substantial fraction (up
to 50\%) of star mass can be lost during the AGB phase via stellar wind.
If the mass-loss process is strong enough, low-mass stars entering the AGB phase can easily 
approach a critical C-O core mass threshold about $M_{\rm core} \simeq 0.52$~M$_\odot$. 
This is the minimum mass 
for stars to fully complete AGB evolution and experience the so-called 
``thermal-pulsing'' phase \citep[][see also Fig.~\ref{hbmorph} above]{dorman,blocker}. 
Along thermal pulses, stars venture in the region of Mira variables and end up 
through the so-called ``superwind phase'' by quickly ejecting their residual envelope and
originate a PN \citep[see][for an exhaustive discussion of the process and its variants]{iben}.

Models indicate that the lack of a full AGB deployment 
(when $M_{\rm core} \la 0.52~M_\odot$) leads to a range of post-HB evolutionary 
paths (see, again, the sketch in Fig.~\ref{hbmorph}),\footnote{From the physical point 
of view, this would correspond to the He+H double-shell burning regime for low- and 
intermediate-mass stars.} as discussed in detail by \citet{greggio}. 
One relevant case in this regard is that of EHB objects, that evolve as 
{\it ``AGB-manqu\'e''} stars, thus fully escaping the PN ejection and fading directly along 
the high-temperature white-dwarf cooling sequence 
\citep{castellani92,castellani,dorman,dcruz96,yi}.

Quite interestingly, therefore, the successful detection of PNe,  even in distant unresolved galaxies,
places a further interesting constraint to the final mass of the composing stellar population.
More specifically, the PN number density per unit galaxy luminosity, a parameter
often referred to in the literature as the ``$\alpha$ ratio'' \citep{jacoby}, can directly 
be linked to the characteristic lifetime of the nebulae, the latter closely tracing the 
mass distribution of their nuclei \citep[see][for a full discussion]{buzzoni06}.
On the basis of these arguments, PNe can eventually help constraining the initial-to-final 
mass relation (IFMR) also in extragalactic systems.

\begin{figure}[!t]
\centerline{
\psfig{file=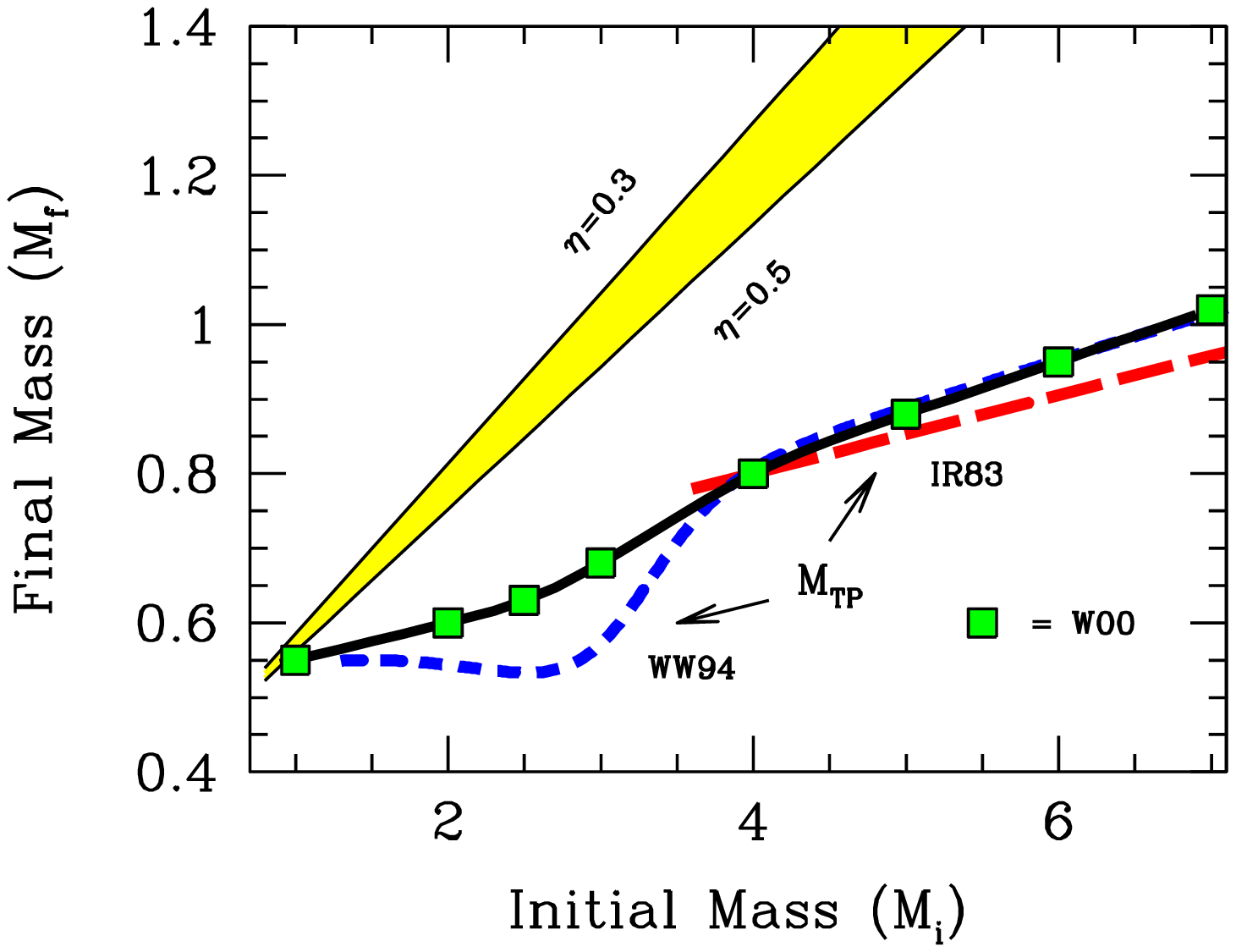,width=0.70\hsize,clip=}
}
\caption{The initial-to-final mass relation according to different
calibrations.  The solid strip is the theoretical relation of
\citet{iben} for a standard mass loss parameter $\eta$ in the
range between 0.3 and 0.5, as labelled on the plot. Short- and
long-dashed curves are the theoretical loci for stars to set on the
AGB thermal pulsing phase ($M_{\rm TP}$), according to \citet{iben} (IR83) and
\citet{ww94} (WW94). Finally, big squares and solid curve report the
\citet{weidemann} (W00) empirical relation based on the mass estimate
of white dwarfs in Galactic open cluster.}
\label{mfin}
\end{figure}

For our galaxy, the IFMR can be derived empirically from the observation of the 
white dwarfs in nearby open clusters (like Hyades or Praesepe) of known age
(as obtained, for instance by the isochrone fitting method of the cluster c-m diagram). 
In an exhaustive study of the available observed database, \citet{weidemann} found evidence 
that white-dwarf masses, for low- and intermediate-mass 
stars, closely match the theoretical core masses expected at the beginning of the thermal-pulsing AGB.
This claim is accounted for in Fig.~\ref{mfin}, where we compare the \citet{weidemann} 
IFMR with the \citet{ww94} updated set of AGB stellar tracks for Pop I stars, and with the 
original analytical relation for thermal-pulsing core mass for intermediate-mass stars by \citet{iben}.
It is clear from Fig.~\ref{mfin} that, for a standard range of the \citet{reimers} mass-loss 
parameters pertinent to Galactic globular clusters 
\citep[namely $\eta \simeq 0.4 \pm 0.1$, according to][]{fusipecci} the
\citet{iben} theoretical IFMR predicts unreliably high final masses for 
young ($t \la 2$~Gyr) SSPs, requiring a value of $\eta \gg 1$ to match the 
\citet{weidemann} empirical relation.

\begin{figure}[!t]
\centerline{
\psfig{file=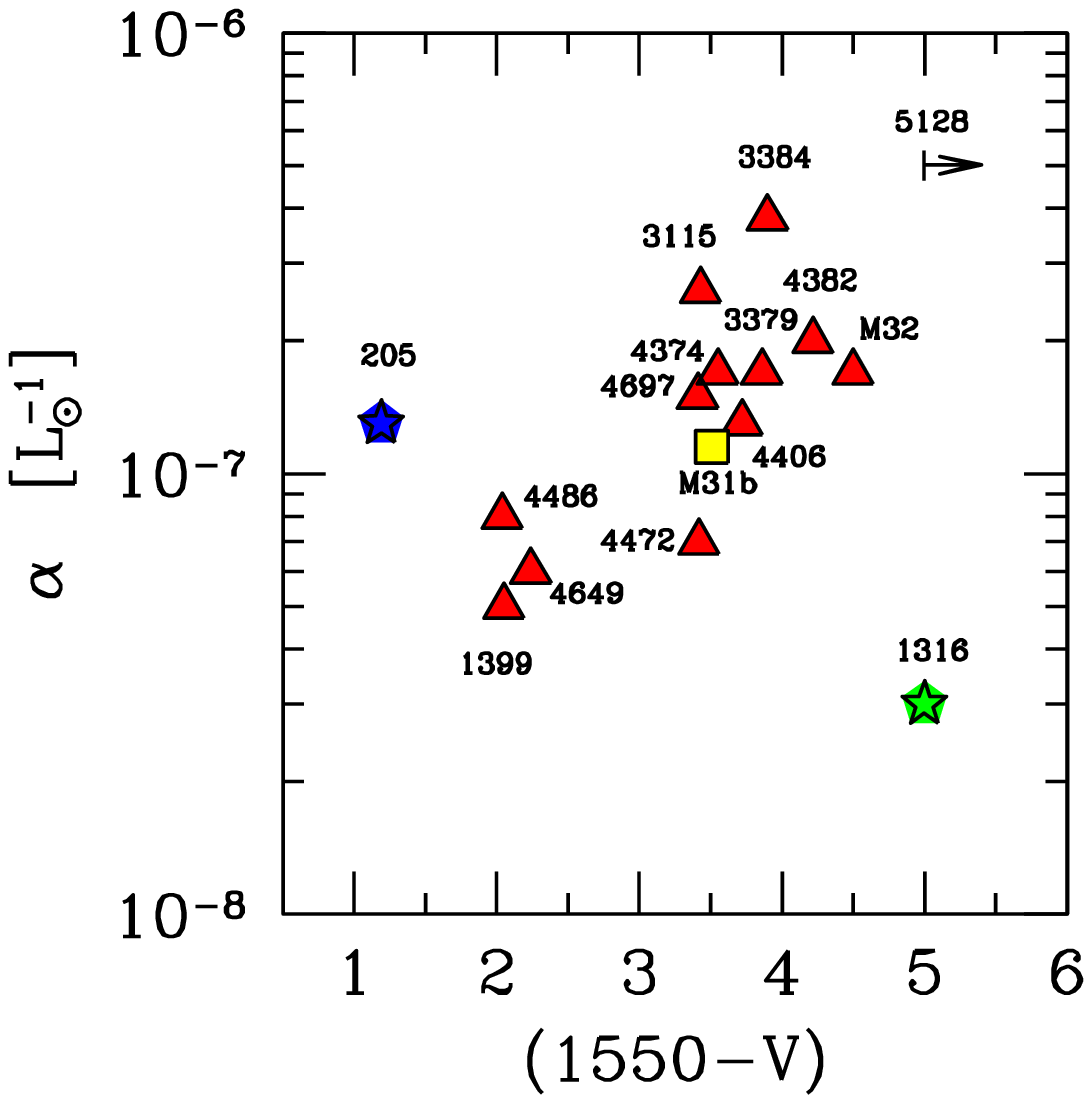,width=0.53\hsize,clip=}
\psfig{file=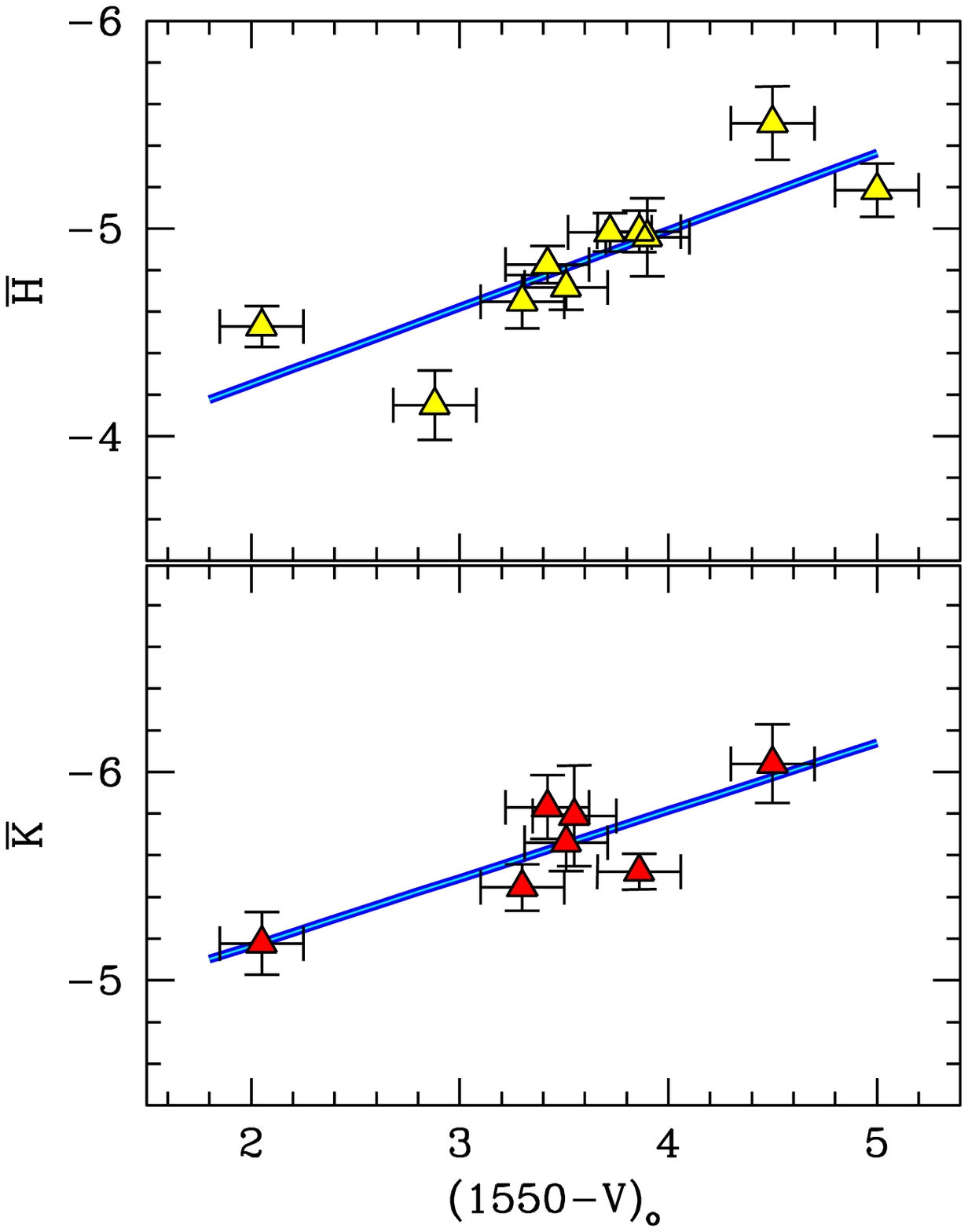,width=0.47\hsize,clip=}
}
\caption{{\it Left panel:} the observed relationship between PN luminosity-specific rate 
$\alpha$ and integrated ultraviolet-to-optical emission (as defined in Fig.~\ref{oconnell})
for a sample of elliptical galaxies in the Local Group and in the Virgo, Leo and Fornax
clusters according to \citet{buzzoni06}. The relevant case of outliers like NGC~205 (star-forming) 
and NGC~1316 (merger elliptical) are singled out in the plot. {\it Right panel:} the corresponding 
relationship in place among the \citet{cantiello03} sample of quiescent ellipticals vs.\
\citet{tsch} effective magnitudes in the H and K infared bands.
The most UV-enhanced galaxies display the faintest infrared effective magnitudes,
sensitive of a less deployed AGB.
}
\label{alfauv}
\end{figure}

As far as external galaxies are concerned, the Local Group represents a natural benchmark 
to assess the IFMR through the estimated $\alpha$ ratio from deep PN surveys.
Quite surprisingly, in spite of the extreme variety of star formation histories among local
galaxies, \citet{buzzoni06} have demonstrated that observations support a fairly constant 
value of $\alpha$, with an average PN rate per unit galaxy luminosity between 1 and 6~PNe per
$10^7$~L$_{\odot}$ among systems representative of the whole late-type Hubble morphological sequence.
Such a value is related to a quite narrow range of final stellar masses about 0.60-0.65~$M_\odot$.

Even in case of Local Group member galaxies, therefore, the mass-loss scenario supported by 
the PN observations better agrees with the \citet{weidemann} IFMR, which implies a substantially 
stronger mass loss for intermediate and high-mass stars compared to the standard scenario 
for Pop II stars as in Galactic globular clusters.

\section{UV evolution and AGB connection in elliptical galaxies}

One important consequence of the {\it ``AGB-manqu\'e''} evolution of EHB stars
is that a tight and {\it inverse} relationship must be in place, especially involving
elliptical galaxies, between the most PN-poor and UV-enhanced stellar systems.
A prevailing fraction of hot (low-mass) HB stars in the galaxy stellar population, in fact, can 
be strongly favoured by a more efficient mass loss (at least along the RGB evolution), 
and this scenario, by itself, also plays against any full AGB deployment.
Therefore, if this is the case, few stars (if any) can eventually reach the AGB feeding 
the PN production channel.

As shown in Fig.~\ref{alfauv} (left panel) this correlation is actually displayed between
PN luminosity-specific rate $\alpha$ and the $(1550-V)$ color for elliptical galaxies in 
the Virgo and Fornax clusters, and in the Leo group, after \citet{buzzoni06}. The sense is that 
more massive metal-rich systems (traced by a higher velocity dispersion $\sigma_v$ and a stronger 
integrated Lick Mg$_2$ index) display at a time a stronger UV-upturn
{\it and} a poorer PN population per unit galaxy luminosity.
The \citet{buzzoni06} relationship settles an old (and so far unexplained) empirical evidence 
for a trend of PN rate, seen to decrease among the reddest ellipticals \citep{hui}.

As $\alpha$ can actually be considered as an indirect probe of galaxy AGB extension above the 
thermal-pulses threshold, this would naturally lead to expect some correlation between the PN 
luminosity-specific rate and galaxy infrared colors or, even better, with more sensitive  
and unbiased tracers of the cool galaxy stellar component. This is, for instance, the case of the 
infrared effective magnitudes, as derived from the surface-brightness fluctuation method of 
\citet{tsch}.
Again, our guess is fully confirmed by a study of the Virgo and Leo elliptical sample 
(Buzzoni \& Gonz\'alez-L\'opezlira, in preparation), as shown in the right panel of 
Fig.~\ref{alfauv} based on \citet{cantiello03} compilation database.

\end{document}